-------------------------------------------------------------------------------
\documentstyle[12pt,twoside,fleqn]{article}
\begin{document}
\setlength{\baselineskip}{5ex}

%
\newcommand{\fns}{\footnotesize}
\newcommand{\bect}{\begin{center}}       \newcommand{\enct}{\end{center}}
\newcommand{\befr}{\begin{flushright}}   \newcommand{\enfr}{\end{flushright}}
\newcommand{\befl}{\begin{flushleft}}    \newcommand{\enfl}{\end{flushleft}}
\newcommand{\beeq}{\begin{equation}}     \newcommand{\eneq}{\end{equation}}
\newcommand{\bear}{\begin{array}}        \newcommand{\enar}{\end{array}}
\newcommand{\bea}{\begin{eqnarray}}      \newcommand{\eea}{\end{eqnarray}}
\newcommand{\beitm}{\begin{itemize}}     \newcommand{\enitm}{\end{itemize}}
\newcommand{\nid}{\noindent}             \newcommand{\hsp}{\hspace*{8mm}}
\newcommand{\hspi}{\hspace*{\parindent}}
\newcommand{\bsk}{\bigskip}              \newcommand{\msk}{\medskip}
\newcommand{\ssk}{\smallskip}
\newcommand{\dps}{\displaystyle}
\newcommand{\ra}{\mbox{$\rightarrow$}}
\newcommand{\Lra}{\mbox{$\Longrightarrow$}}
\newcommand{\ctl}[1]{\centerline{#1}}
\newcommand{\frlr}[3]{\left(\frac{#1}{#2}\right)^{#3}} 
\newcommand{\frlrsq}[2]{\left(\frac{#1}{#2}\right)^2}  
\newcommand{\braces}[1]{\left\{#1\right\}}  
\newcommand{\Lbr}[1]{\left[#1\right]}       
\newcommand{\Lprth}[1]{\left(#1\right)}     

\newcommand{\ie}{{\it i.\,e.\ }}
\newcommand{\eg}{{\it e.\,g.\ }}
\newcommand{\etal}{{\it et al.}}

\newcommand{\al}{\mbox{$\alpha$}}
\newcommand{\bt}{\mbox{$\beta$}}
\newcommand{\gm}{\mbox{$\gamma$}}
\newcommand{\Gm}{\mbox{$\Gamma$}}
\newcommand{\dl}{\mbox{$\delta$}}
\newcommand{\Dl}{\mbox{$\Delta$}}
\newcommand{\ep}{\mbox{$\epsilon$}}
\newcommand{\vrep}{\mbox{$\varepsilon$}}
\newcommand{\vep}[1]{\mbox{$\varepsilon_{#1}$}}  
\newcommand{\zt}{\mbox{$\zeta $}}
\newcommand{\th}{\mbox{$\theta$}}
\newcommand{\kp}{\kappa}
\newcommand{\ld}{\mbox{$\lambda$}}
\newcommand{\Ld}{\mbox{$\Lambda$}}
\newcommand{\vp}{\mbox{$\varpi$}}           
\newcommand{\vre}{\mbox{$\varrho_{\rm e}$}} 
\newcommand{\vrm}{\mbox{$\varrho_{\rm m}$}} 
\newcommand{\sg}{\mbox{$\sigma$}}
\newcommand{\Sg}{\mbox{$\Sigma$}}
\newcommand{\vh}{\mbox{$\varphi$}}
\newcommand{\om}{\mbox{$\omega$}}
\newcommand{\Om}{\mbox{$\Omega$}}
\newcommand{\OmH}{\mbox{$\Omega_{\rm H}$}}  
\newcommand{\OmF}{\mbox{$\Omega_{\rm F}$}}  
\newcommand{\OmS}{\Omega_{\rm S}}           
\newcommand{\nb}{\mbox{$\nabla$}}

\newcommand{\vc}[1]{\mbox{\boldmath $#1$}}    
\newcommand{\vcnbl}{\vc{\nb}}                 
\newcommand{\pl}{\partial}
\newcommand{\divg}[1]{\vcnb\cdot\vc{#1}}       
\newcommand{\rot}[1]{\vcnb\times\vc{#1}}       
\newcommand{\rotal}[1]{\vcnb\times\al\vc{#1}}  
\newcommand{\grad}[1]{\vcnb#1}                 
\newcommand{\dr}[2]{\frac{d#1}{d#2}}           
\newcommand{\drh}[3]{\frac{d^{#1}#2}{d#3^{#1}}}        
\newcommand{\pldr}[2]{\frac{\pl#1}{\pl#2}}     
\newcommand{\pldh}[3]{\pl^{#1}_{#2}#3}         
\newcommand{\pldrh}[3]{\frac{\pl^{#1}#2}{\pl#3^{#1}}}  
\newcommand{\pldrhfp}[4]{\Lprth{\frac{\dps\pl^{#1}#2}{\dps\pl#3^{#1}}}_{#4}}
\newcommand{\pldrhfbr}[4]{\Lbr{\frac{\pl^{#1}#2}{\pl#3^{#1}}}_{#4}}
\newcommand{\pldrt}[2]{\pl#1/\pl#2}             
\newcommand{\gradpl}[2]{(\vc{#1}\cdot\vcnb)#2} 
\newcommand{\vcpr}[2]{\vc{#1}\times\vc{#2}}    

\newcommand{\bh}{black hole}
\newcommand{\ns}{neutron star}

\newcommand{\ApJ}{{\it Astrophys.\,J.}}
\newcommand{\CQG}{{\it Class.\,Quantum Grav.}}
\newcommand{\MN}{{\it Mon.\,Not.\,R.\,Astron.\,Soc.}}
\newcommand{\PASJ}{{\it Publ.\,Astron.\,Soc.\,Japan}}
\newcommand{\PRD}{{\it Phys.\,Rev.\,}D}
\newcommand{\PL}{{\it Phys.\,Lett.}}
\newcommand{\GRG}{{\it Gen.\,Rel.\,Grav.}}
\newcommand{\PRSL}{{\it Proc.\,Roy.\,Soc.\,Lond.}}
\newcommand{\ibid}{{\it ibid}}
\newcommand{\vol}[1]{{\bf #1}}        


\newcommand{\NV}[2]{#1\cdot 10^{#2}}
\newcommand{\plx}{\partial_{x}}          \newcommand{\plE}{\partial_{E}}
\newcommand{\qmf}[1]{\overline{(#1)^2}}    
\newcommand{\Dlx}{\Delta x}
\newcommand{\tlS}{\tilde{S}}                 
\newcommand{\tlbt}{\tilde{\beta}}        \newcommand{\tlP}{\tilde{P}}
\newcommand{\tlH}{\tilde{H}}
\newcommand{\Dlbt}{\Dl\tlbt}             \newcommand{\qmfDbt}{\qmf{\Dlbt}}
\newcommand{\tlzt}{\tilde{\zeta}}        \newcommand{\Dlzt}{\Dl\tlzt}
\newcommand{\qmfDzt}{\qmf{\Dlzt}}
\newcommand{\btC}{\bt_{\rm c}}           \newcommand{\EC}{E_{\rm c}}
\newcommand{\Srad}{S_{\rm rad}}   \newcommand{\Sbhrad}{S_{{\rm bh}+{\rm rad}}}
\newcommand{\EB}{E_{{\rm B}}}          \newcommand{\Ebh}{E_{{\rm bh}}}
\newcommand{\Erad}{E_{{\rm rad}}}        \newcommand{\btrad}{\bt_{{\rm rad}}}
\newcommand{\Cbh}{C_{{\rm bh}}}          \newcommand{\Crad}{C_{{\rm rad}}}
\newcommand{\qmfbtrad}{\qmf{\Dlbt_{{\rm rad}}}}
\newcommand{\qmfbtbh}{\qmf{\Dlbt_{{\rm bh}}}}
\newcommand{\Pc}{Poincar\'{e}}
\newcommand{\Schw}{Schwarzschild}        \newcommand{\KN}{Kerr-Newman}


\befl
   {\normalsize {\bf A comment on fluctuations and stability limits with
   application to ``superheated" black holes } }
\enfl
\msk
\renewcommand{\thefootnote}{\fnsymbol{footnote}}
\small
\footnotetext[1]{Permanent address: Division of Theoretical Astrophysics,
      National Astronomical Observatory, Mizusawa, Iwate 023, Japan;
      okamoto@gprx.miz.nao.ac.jp}
\footnotetext[2]{jkatz@hujivms.bitnet}
\footnotetext[3]{present address:  Laboratoire de Physique Theorique de l'
      Ecole Normale Superieure, 24 Rue Lhomond, 75005 Paris, France (E-mail:
      parenta@physique.ens.fr)}
\befr \parbox[t]{14.4cm}{ %
 {\bf Isao Okamoto\footnotemark[1], Joseph Katz\footnotemark[2], and
      Renaud Parentani\footnotemark[3]}\\
      The Racah Institute of Physics, 91904 Jerusalem, Israel \\[5mm]
      Received %
}
\enfr
\msk
\befr\parbox[t]{14.4cm}{ %
{\bf Abstract.} We point out that, contrary to signs of heat capacities,
thermodynamic fluctuations are simply and unequivocally related to onset of
instabilities that show up near critical points. Fluctuation theory is then
applied to \Schw\ \bh s surrounded by radiation. This shows that slowly
evolving \bh s  along quasi-equilibrium states in cavities greater than $10^6$
Planck length will not evaporate below the critical Hawking limit temperature
despite the fact that pure radiation has a much higher entropy.
   } \\
\enfr
\bsk
In a mean field approximation, self-gravitating thermodynamic ensembles have
local maxima of entropy or other Massieu functions derived from the entropy
by a Legendre transformation, even though the potential energy is not bounded
from below (see \eg Horwitz and Katz 1978).  These ensembles are generally not
equivalent. This means that the domains of stability of different ensembles
related by a Legendre transformation are different. Arguments based on signs
of heat capacities and similar thermodynamic coefficients are dubious criteria
of stability. Far more secure results are derived from linear series (\ie
equilibrium curves) using \Pc's theorem about bifurcations or turning points
(Ledoux 1958). One of the most elegant ways of using \Pc's method consists in
following equilibrium curves of pairs of conjugate parameters: inverse
temperature versus energy, angular velocity versus angular momentum, and so
on.\ (Katz 1978, 1979).  This is the method used in Kaburaki \etal\ (1993) for
Kerr \bh s and in Katz \etal\ (1993) for \KN\ \bh s to find stability limits
of stable configurations and degrees of instability in unstable configurations
[see Parentani (1994) for the origin of the inequivalence and the degree of
stability in canonical ensembles].

Here we want to stress that fluctuation theory is closely related to linear
series of conjugate parameters and is indeed applicable to thermodynamic
ensembles whether they are equivalent or not\footnotemark[4].
\footnotetext[4]{See note added in proof}
This is readily seen as follows.

Think, for definition, of a \Schw\ \bh\ in a sphere filled with photons at the
same temperature (Hawking 1976). The total entropy of the system is $S=S(E,V)$,
 where $E$ is the total energy of hole and radiation and $V$ is the volume of
the cavity.  Conjugate pairs with respect to $S$ are $(\bt,E)$ and $(P,V)$
with
\beeq
  \bt=\pldr{S}{E}, \qquad P=\frac{1}{\bt}\pldr{S}{V}.
     \eneq  
Here $\bt^{-1}$ is the temperature, $P$ the pressure (units are $c\!=\!\hbar\!
=\!G\!=\!k\!=\!1$). Linear series of conjugate variables are the equilibrium
curves $\bt(E)$ at fixed $V$ (see figure 1) and $P(V)$ at fixed $E$ (not
drawn). Changes of stability in the microcanonical ensemble (\ie at fixed $E$
and $V$) occur only at the critical point C, where there is a vertical tangent
($(\pl\bt/\pl E)_V=\pm\infty$). Positive slopes {\em near} C are on a branch
of stable configurations, while negative slopes near C on one of unstable
configurations.  The proof is general and has been given elsewhere but is
quite useful to remember and short enough to be reproduced here for the cases
in which there is only one fluctuating variable $x$. This is the situation of
the Hawking example where $x$ is the partition energy $\Ebh$, the energy of
the black hole and where the left over energy of the radiation is  $\Erad=E-
\Ebh$.

Let $\tlS(x;E,V)$ be the entropy away from equilibrium. Equilibrium is defined
by extremising the entropy with respect to $x$
\beeq
  \plx\tlS=0
     \eneq  
giving two solutions $x=X(E,V)$ and $x=X_1(E,V)$. Denote $(X,X_1)$ collectively
 by $X_a$ ($a=$nothing or $1$) and $\tlS(X_a;E,V)=S_a(E,V)$.  It is useful to
{\em define} a temperature away from equilibrium
\beeq
   \tlbt=\plE\tlS=\tlbt(x;E,V).
      \eneq 
Clearly this function evaluated along the equilibrium configurations $x=X_a$
gives
\beeq
   (\tlbt)_{x=X_a}=\bt_a=\pldrhfp{}{S_a}{E}{V}
      \eneq 
which is the usual equilibrium inverse temperature. The equilibrium curves
$\bt(E)$ and $\bt_1(E)$ are represented in figure 1.

Stable equilibria exist if and only if $x=X_a$ is a configuration at which
$\tlS$ is maximum:
\beeq
  \Lprth{\plx^2\tlS}_a<0.
     \eneq  
Consider now equation (2) in which $x$ is replaced by $X_a$; (2) reduces then
to $0=0$ and the following identity holds
\beeq
  \pldrhfbr{}{(\plx\tlS)_a}{E}{V}
     =(\plx\tlbt)_a+(\plx^2\tlS)_a\pldrhfp{}{X_a}{E}{V}\equiv 0.
       \eneq 
On the other hand, a further derivative of (4) with respect to $E$ with $V$
kept fixed gives
\beeq
   \pldrhfp{}{\bt_a}{E}{V}=\Lprth{\plE^2\tlS }_a
     +(\plx\tlbt)_a\pldrhfp{}{X_a}{E}{V}.
        \eneq 
Eliminating $\pldrt{X_a}{E}$ between (6) and (7) gives then
\beeq
   \pldrhfp{}{\bt_a}{E}{V}=\Lprth{\plE^2\tlS}_a
     -\frac{(\plx\tlbt)_a^2}{(\plx^2\tlS)_a}.
        \eneq 

Since changes of stability occur when $(\plx^2\tlS)_a$ goes through zero, they
manifest themselves by vertical slopes because for $(\plx^2\tlS)_a\ra 0$
\beeq
   \pldrhfp{}{\bt_a}{E}{V}\approx -\frac{(\plx\tlbt)_a^2}{(\plx^2\tlS)_a}
      \to \pm\infty
         \eneq 
(see figure 1). The point of a vertical slope with coordinates $(\EC,\btC)$ is
a critical point. Near a critical point the slope is positive on the stable
branch $\bt(E)$ where $(\plx^2\tlS)_a<0$  and negative on the unstable branch
$\bt_1(E)$.

Consider now $\tlS(x;E,V)$ near a point of stable equilibrium, at $x=X+\Dlx$:
\beeq
   \tlS=S+\frac12(\plx^2\tlS)(\Dlx)^2+O\Lbr{(\Dlx)^3}
         \eneq 
or, according to (9)
\beeq
    \tlS \approx
         S-\frac12\frac{(\plx\tlbt)^2(\Dlx)^2}{\pldrhfp{}{\bt}{\dps E}{V}}
         =S-\frac12\frac{(\Dl\tlbt)^2}{\pldrhfp{}{\bt}{\dps E}{V}}
            \eneq 
where $\Dl\tlbt=\tlbt-\bt$ and
\beeq 
  \pldrhfp{}{\bt}{E}{V}>0.
      \eneq
$\Dl\tlbt$ is a deviation $\tlbt(x)$ from stable equilibrium values $\bt(E)$
induced by the fluctuation of $x$ away from the equilibrium configuration
$X(E,V)$. The probability $dW$ of such a fluctuation of $\tlbt(x)$ in a range
$(\tlbt+\Dl\tlbt, \tlbt+\Dl\tlbt+d\tlbt)$ is proportional to $\exp\,(\tlS-S)$
[see Landau and Lifshitz (1980)]. The normalization factor of $dW$ is easy to
find in this quadratic approximation:
\beeq
   dW=\frac{1}{\sqrt{2\pi\qmfDbt}}
       \exp\Lbr{-\frac12\frac{(\Dl\tlbt)^2}{\qmfDbt}}d\tlbt
            \eneq 
where
\beeq
    \qmfDbt=\pldrhfp{}{\bt}{E}{V}.
         \eneq 
The mean square fluctuation $\qmfDbt$ of $\tlbt$ is thus given {\em near the
critical point} by the positive slope of $\bt(E,V)$, \ie where the specific
heat is negative. [Note however that in the canonical ensemble, at fixed $\bt$,
 the mean square fluctuations of the total energy are given by the slope of
$-E(\bt,V)$ as usual (see (19.6) in Callen 1985).] The real physical mean
square fluctuations of temperatures in the bath of photons $\qmfbtrad$ and in
the \bh\ $\qmfbtbh$ are neither equal to $\qmfDbt$ nor equal to each other
because subsystems have different heat capacities. It is, however, not hard to
show that near the critical point
\beeq
     \lim_{x\to X_a}\frac{\qmfbtrad}{\qmfDbt}
    =\lim_{x\to X_a}\frac{\qmfbtbh}{\qmfDbt}=1.
          \eneq 
Thus $\qmfDbt$ as given in (14) or (9) represents, to a good degree of
approximation, the quadratic fluctuations of temperature that diverge at a
critical point. Similar fluctuations of a pressure function $\tlP(x;E,V)=
\bt^{-1}\pl_V\tlS$ might have been considered.

The more general relation between positive slopes and mean square
fluctuations of a conjugate variable in linear series near the critical points
where a change of stability takes place is obvious: if $\tlzt$ (like $\tlbt$)
is the conjugate function of $Z$ which, like $E$, is a control parameter with
respect to some Massieu function $\tlH(\tlzt;Z)$ (similar to $\tlS(x;E,V)$),
the equilibrium value of $\tlzt$ is
\beeq
    \zt=\pldr{H}{Z}.
         \eneq 
We assume that second derivatives of $\tlH$ are bounded and that, at the
critical point, $(\plx\tlH)_a\neq 0$. If $(\plx\tlH)_a=0$, vertical slopes
may be transformed into bifurcations. But very small perturbation of $\tlH$
like $\tlH+\vrep xE$, $|\vrep|\ll|xE/\tlH|$ will easily transform bifurcation
into turning points (Thompson and Hut 1973) with almost no change in stability
limits.

In these circumstances, near the critical point, the probability of $\tlzt$
fluctuating by $\Dl\tlzt=\tlzt-\zt_{\rm c}$ is given by
\beeq
   dW=\frac{1}{\sqrt{2\pi\qmfDzt}}
       \exp\Lbr{-\frac12\frac{(\Dl\tlzt)^2}{\qmfDzt} }d\tlzt,
           \quad
              \qmfDzt=\pldr{\zt}{Z}>0.
                  \eneq 

When there are more variables $x$, we imagine that the $x^i$'s have been taken
in such a way that $-(\pl_{ij}^2\tlH)_a$ is diagonal, say $\dl_{ij}\ld_j$. The
$\ld_j$ are called the \Pc\ coefficients of stability.  If the spectrum of
$\ld_i$'s is non-degenerate (which is always possible to achieve with a small
perturbation of $\tlH$), there will be a vertical slope at each point where a
change of stability occurs and at the particular point $\pldrt{\zt}{Z}$ will
be the same approximate value given by (9).

For a simple application of fluctuation theory, we go back to (13) in the case
of the \Schw\ \bh\ in a cavity for which $\bt(E)$, given in Hawking (1976)
and Gibbons and Perry (1978), is
\beeq
    E=\frac{\bt}{8\pi}+\frac{\pi^2}{15}V\bt^{-4}, \quad
        V=\frac{4\pi}{3}L^3
            \eneq 
where $L$ is the radius in Planck length unit. Suppose the \bh\ evolves
through a series of quasi-equilibrium configurations by extracting slowly
energy out of the box whose size is kept fixed. We know that at the critical
point the hole will certainly evaporate. What we want to estimate is the
probability per unit time that the \bh\ evapolates before reaching its
stability limit as a result of thermal fluctuations.

{}From (14) and (18)
\beeq
    \qmfDbt=\pldrhfp{}{\bt}{E}{V}
         =\frac{8\pi}{\Lbr{1-\Lprth{\frac{\dps\btC}{\bt}}^5}}
              \eneq 
in which
\beeq
    \btC=\Lprth{\frac{128\pi^4}{45}}^{\frac15}L^{\frac35}
         \approx 3.1 L^{\frac35}.
               \eneq 
The probability of complete evaporation of the \bh\ is essentially the
probability of the system fluctuating from $\bt(E)$ to $\bt_1(E)$. Indeed,
as soon as $\tlbt$ is slightly bigger than $\bt_1$, the probability of going
back to $\bt$ is negligible for $\Srad-S_1\gg S-S_1$. Therefore, to estimate
$dW$, let us take for a typical value of $\Dl\tlbt$ the following one
$\Dlbt=\bt-
\btC$ (taking $\bt-\bt_1$ would give an underestimate of that probability
because the quadratic approximation is no longer valid over such an interval).
With $|(\bt-\btC)/\btC|\ll 1$ and with (18) and (19), $dW$ as given in (13)
becomes
\beeq 
   dW\Lprth{\frac{\Dlbt}{\btC},L}
      \approx 0.2\frlr{\Dlbt}{\btC}{\frac12}
         \exp\Lbr{-0.9\frlr{\Dlbt}{\btC}{3}L^{\frac65}}d\tlbt.
             \eneq
Hence, for a given value of $\Dlbt/\btC$, the probability of evaporation
depends only on the size of the box. For instance, say, $\Dlbt/\btC=0.1$
and $L=10^3$, $dW=10^{-3}d\tlbt$ but for $L=10^4$, $dW=\NV{8}{-28}d\tlbt$, \ie
$dW$ drops drastically for $L>10^3$. If $\Dlbt/\btC=0.01$ and $L=10^6$,
$dW=\NV{6}{-9}d\tlbt$ but for $L=10^7$, $dW\simeq \NV{2}{-105}d\tlbt$.

Having found $dW$, we may now estimate the rate of evaporation. When the
equilibrium configuration evolves slowly from P to C (see figure 1), the
characteristic time scale of the return to equilibrium after a jump in
energy of $\Dlx$ is of the order of $\Dl t\approx \bt^4\Dlx$ (see Zurek
1980). Since the characteristic time $t_{{\rm c}}$ for a jump from point P
at $\bt(E)$ to point Q at $\bt_1(E)$ is also proportional to $\bt^4$, it
follows that the probability of evapolation per unit time is of the order of
$dW/\bt^4(\btC-\bt)$ (see Piran and Wald 1982 for a similar argument).

What the slope of $\bt(E)$ shows therefore is that for $L>10^6$, fluctuations
of temperature leading to complete evapolation are totally negligible and the
evolution of the \bh\ proceeds down almost to point C before evaporation takes
place. The bigger the box, the closer to C.  This holds in spite of the fact
that near C, configurations are metastable and evaporated \bh s (\ie pure
radiation) have considerably more entropy: $\Srad-\Sbhrad\approx L^{\frac65}
>10^7$.

Let us add that for $L>10^6$, backreaction and quantum gravity are totally
negligible effects (York 1985) and that the radiation is far from being
general relativistic near the critical point (negligible self-gravity\,---\,see
 Parentani \etal\ 1994). Thus, Hawking's approximation for $S$ on which our
approximations are based is perfectly good. Instead, for smaller boxes, the
thermodynamic analysis probably loses its validity because the quantum
fluctuations become important since one approaches the Planck temperature at
point C. \bsk

\befl {\bf Acknowledgement} \ssk \enfl

\nid
IO thanks A Shapere from Cornell University for useful conversation at the
Racah Institute on December 1992. \bsk

\nid
{\it Note added in proof}. Since this work has been submitted we have done
further work (Parentani \etal\ 1994) on the inequivalence of ensembles when
long-range forces are present. A general analysis of the behavior of the
fluctuations is presented. \bsk

\befl {\bf References} \enfl
\ssk
Callen H B 1985 {\it Thermodynamics and introduction to thermostatics}
p425 (Wiley: New York) \\
Gibbons G W and Perry M J 1978 \PRSL\ \vol{A358} 467 \\
Hawking S W 1976 \PRD\ {\bf 13} 191 \\
Horwitz G and Katz J 1978 \ApJ\ \vol{222}, 941 \\
Kaburaki O, Okamoto I and Katz J 1993 \PRD\ \vol{47} 2234 \\
Katz J 1978 \MN\ \vol{183} 765 \\
------\,1979 \MN\ \vol{189} 817 \\
Katz J, Okamoto I and Kaburaki O 1993 \CQG\ \vol{10} 1323 \\
Landau L D and Lifshitz E M 1980 {\it Statistical physics} 3rd ed \S 110
  (Pergamon: Oxford) \\
Ledoux P 1958 {\it Handbuch Phys.}\ \vol{51} 605 \\
Parentani R 1994 ``The equivalence of thermodynamic ensembles" Preprint
gr-qc 9410017\\
Parentani R, Katz J and I Okamoto 1994 ``Thermodynamics of a black hole
in a cavity" Preprint gr-qc 9410015 (submitted to \CQG) \\
Piran T and Wald R M 1982 \PL\ \vol{90A} 20 \\
Thompson J M T and Hut G W 1973 {\it A general theory of elastic stability}
(Wiley: New York) \\
York J W Jr.\ 1985 \PRD\ \vol{31} 775 \\
Zurek W H 1980 \PL\ \vol{77A} 399 \\

\newpage

\befl
{\bf Figure caption } \\[5mm]
\enfl
{\bf Figure 1.}  Qualitative features of Hawking's $\bt(E)$, $\bt_1(E)$ and of
$\btrad(E)$. $\Dl\tlbt(P)$ is the fluctuation around a point of equilibrium P.
 The critical fluctuation which will cause the evaporation of \bh\ are of the
order of $\bt(P)-\bt_1(Q)$. The vertical line $E=\EB(V)$ is that for which
$S=\Srad(E)$. For $E<\EB$ the entropy $S(E)$ is smaller than $\Srad(E)$ at
fixed volume $V$. Thus, any point P between C and B represents a superheated
metastable (in fact a stable\,--\,see the text) \bh\ in equilibrium with
radiation. Correspondingly, the portion of $E>\EB$ on the curve of $\btrad(E)$
expresses Gibbons and Perry's (1978) superheated radiation state.

\end{document}